# The Transversity Function and Double Spin Azimuthal Asymmetry in Semi-Inclusive Pion Leptoproduction

Elvio Di Salvo

Dipartimento di Fisica and I.N.F.N. - Sez. Genova, Via Dodecaneso, 33

- 16146 Genova, Italy

## Abstract

We show that the transverse momentum dependent transversity function is proportional to the longitudinal polarization of a quark in a transversely polarized proton. This result suggests an alternative, convenient method for determining transversity, without knowing unusual fragmentation functions. The method consists of measuring the double spin azimuthal asymmetry in semi-inclusive pion leptoproduction by a transversely polarized proton target. The asymmetry, which is twist 3, is estimated to be more than 10% under the most favorable conditions. The experiment we suggest is feasible at facilities like DESY and CERN.

# 1 Introduction

The transversity function, usually denoted as $h_1(x)$, is one of the three leading twist distributions of the quark inside the nucleon, together with the unpolarized distribution, $q(x)$, and with the longitudinally polarized one, $\Delta q(x)$. In an infinite momentum frame where the nucleon spin is perpendicular to its momentum, one has

$$h_1(x) = q_\uparrow(x) - q_\downarrow(x), \qquad (1)$$

where $q_{\uparrow(\downarrow)}(x)$ is the probability density to find a quark whose spin is parallel (opposite) to the nucleon spin. Important information on quark dynamics inside the nucleon can be extracted from transversity. For example, the difference $h_1(x) - \Delta q(x)$ is quite sensitive to the orbital angular momentum of the quarks. This is why high energy spin physicists have been concentrating their efforts for several years in determining transversity[1-5]. However, this appears a particularly difficult task[6-10], owing to the chiral odd character of $h_1$, unlike $q$ and $\Delta q$. Different observables sensitive to transversity have been singled out. Among them we recall the Drell-Yan (DY) double spin asymmetry [1, 9] and the single [7, 11, 12, 13] and double [14, 15] spin asymmetry in semi-inclusive deep inelastic scattering (SIDIS).

For the moment SIDIS reactions - realized or planned by HERMES[11, 12], SMC[13] and COMPASS collaborations - appear most promising for determining $h_1$. These reactions are of the type

$$\ell \vec{p} \to \ell' h X, \qquad (2)$$

where $\ell(\ell')$ is a charged lepton, $\vec{p}$ a polarized proton target and $h$ a hadron. If $h$ is a spinning particle whose polarization can be determined (e. g., a $\Lambda$), reaction (2) is kinematically isomorphic to the double spin DY process. However, as we shall discuss in the following, it is more convenient to detect a pion in the final state. In this case we are faced with a single spin asymmetry, therefore we have to exploit a possible azimuthal asymmetry[4, 16, 17, 18, 19, 20] of reaction (2). Indeed, taking a frame whose origin coincides with the location of the proton target and whose $z$-axis is along the momentum transfer of the lepton, the angular distribution of the pion may present, in principle, an azimuthal asymmetry with respect to the plane passing through the proton spin and through the $z$-axis. Incidentally, such kind of asymmetry may be sensitive to various transverse momentum dependent (t.m.d.) distributions[16, 17, 21, 22, 23, 24], defined as "new" by Kotzinian and Mulders[21] (KM); see also Mulders and Tangerman[16] (MT). In order to extract distributions from azimuthal asymmetries, it is convenient to define weighted asymmetries[19, 21], that is, ratios of the type

$$\langle A_w \rangle = \frac{\int w d\sigma}{\int d\sigma}. \quad (3)$$

Here $w$ is a given function of the azimuthal angle $\phi$ of the final hadron $h$, over which we perform the integrations indicated in formula (3). In the reaction considered the azimuthal asymmetry is caused by the Collins effect[25], consisting of an interference term[26]. This effect is described by a T-odd (and chiral odd) t.m.d. fragmentation function, say $\kappa$. The corresponding weighted asymmetry results to be sensitive to the product $h_1(x)c(z)$, where $c = \int w\kappa d\phi$ is the so-called Collins fragmentation function[25]. As claimed by Jaffe[4], this may become the "classic" way of determining the proton transversity distribution, provided $c(z)$ is known to some precision and is not too small. But at present we know very little about this function, moreover the methods suggested for inferring it from data are complicated and require an adequate statistics[20, 4].

Analogous considerations could be done about the method proposed by Jaffe and Ji[27] (JJ). This consists of measuring the double spin asymmetry in a SIDIS reaction of the type (2), where the initial lepton is longitudinally polarized, the proton target is transversely polarized and the final hadron is a pion, i. e.,

$$\vec{\ell}p^\uparrow \to \ell'\pi X. \quad (4)$$

The corresponding asymmetry is defined as

$$A(|\mathbf{k}|; Q, \nu; \Pi_\parallel) = \frac{d\sigma_{\uparrow\to} - d\sigma_{\uparrow\leftarrow}}{d\sigma_{\uparrow\to} + d\sigma_{\downarrow\leftarrow}}. \quad (5)$$

Here, as usual, $\nu$ is the lepton energy transfer and $Q^2 = -q^2$, $q$ being the four-momentum transfer. Furthermore $\mathbf{k}$ is the momentum of the initial lepton and $\Pi_\parallel$ the component of the pion momentum along the momentum transfer. Lastly $d\sigma_{\uparrow\to}$ and $d\sigma_{\uparrow\leftarrow}$ are the polarized differential cross sections for reaction (4) integrated over the azimuthal angle of the pion, arrows indicating the proton and lepton polarization. Asymmetry (5) is sensitive to the product $h_1(x)\hat{e}(z)$[27], where $\hat{e}(z)$ is the twist-3 fragmentation function of the pion. The extraction of $h_1$ depends again critically on an unknown function.

If the cross section is not integrated over the transverse momentum of the final pion, reaction (4)[21, 17] exhibits an azimuthal asymmetry. This is sensitive[21] to the "new" function $g_{1T}$, proportional to the longitudinal quark polarization in a transversely polarized proton.

The aim of this paper is to re-examine such an azimuthal double spin asymmetry. We derive the differential cross section for reaction (2), starting from the definition of t.m.d. transversity function as given by Jaffe and Ji[2] (JJ1), i. e.,

$$\delta q_\perp(x, \mathbf{p}_\perp) = \sum_{T=\pm 1/2} 2T q_T(x, \mathbf{p}_\perp). \tag{6}$$

Here, analogously to formula (1), $q_T(x, \mathbf{p}_\perp)$ is the probability density to find, in a transversely polarized proton, a quark whose spin is parallel ($T = 1/2$) or opposite ($T = $-1/2) to the proton spin. That is, instead of the usual helicity representation, we consider a canonical one, such that the quantization axis is taken along the proton polarization. The asymmetry we calculate turns out to coincide with the one by KM, provided we identify * $g_{1T}$ with $\delta q_\perp$. We shall prove this identity for massless quarks, which amounts to saying that, owing to transverse momentum, a quark in a transversely polarized proton has a longitudinal polarization, related to transversity. Therefore $\delta q_\perp$ - denoted as $h_{1T}$ by MT - plays a major role in the azimuthal asymmetry of reaction (4), where the target is transversely polarized. Moreover, as we shall see, this distribution is somewhat relevant also in the case of a longitudinally polarized target. All this suggests an alternative, convenient method for determining the transversity. Indeed, as a consequence of our result, $\delta q_\perp$ contributes not only to the chiral-odd component of the t.m.d. correlation matrix[16, 23, 28], but also to its chiral-even part. Therefore this distribution may be coupled to a chiral-even fragmentation function, which is generally easier to determine than a chiral-odd one. This result is not completely surprising, since we have shown in a previous paper[29] that DY from singly polarized proton-hadron collisions - which is kinematically isomorphic to reaction (4) - produces a muon polarization sensitive to $\delta q_\perp$.

---
*up to a normalization constant, see appendix and subsect. 2.4

Drawbacks, common to any SIDIS reaction, have been pointed out recently[30, 26, 31, 32]. In fact, the asymmetries measured in these reactions are likely to derive non-negligible contributions from spurious effects, caused by interactions[30, 26] or correlations[32] between the active quark and the spectator partons, or by fragmentation from the target remnant diquark, or by light cluster decays[31]. Such effects have to be taken into account by means of suitable fracture functions[26].

Sect. 2 is dedicated to the derivation of the formulae for the cross section and for the asymmetry we are interested in. Moreover we illustrate some consequences of the identity $g_{1T} = \delta q_\perp$ in the chiral limit, which we prove in the appendix. In sect. 3 we discuss how to infer the transversity from data of azimuthal double spin asymmetry, taking into account the spurious effects. In sect. 4 we compare our method with others. Lastly sect. 5 is devoted to a short conclusion.

## 2 Cross section and azimuthal asymmetry

### 2.1 Cross section

We calculate the differential cross section for reaction (4) in the framework of a QCD-improved parton model[33]. For the moment we neglect the contribution of the fracture function, which will be discussed in sect. 3. In the laboratory frame the differential cross section reads, in one-photon exchange approximation,

$$d\sigma = \frac{1}{4|\mathbf{k}|M} \frac{e^4}{Q^4} L_{\mu\nu} H^{\mu\nu} \, d\Gamma, \tag{7}$$

where $M$ is the proton rest mass and $L_{\mu\nu}$ ($H_{\mu\nu}$) the leptonic (hadronic) tensor. $d\Gamma$, the phase space element, reads

$$d\Gamma = \frac{1}{(2\pi)^6} d^4k' \, \delta(k'^2) \, \theta(k'_0) \, d^4P \, \delta(P^2 - m_\pi^2) \, \theta(P_0). \tag{8}$$

Here $k'$ and $P$ are, respectively, the four-momenta of the final lepton and of the pion, whose rest mass is $m_\pi$. The leptonic tensor is, in the massless approximation,

$$L_{\mu\nu} = \frac{1}{4}Tr[\slashed{k}(1+\lambda_\ell\gamma_5)\gamma_\mu\slashed{k}'\gamma_\nu], \tag{9}$$

$\lambda_\ell$ and $k = k' + q$ ($k^2 = 0$) being respectively the helicity and the four-momentum of the initial lepton. Trace calculation yields

$$L_{\mu\nu} = k_\mu k'_\nu + k'_\mu k_\nu - g_{\mu\nu} k \cdot k' + i\lambda_\ell \varepsilon_{\alpha\mu\beta\nu} k^\alpha k'^\beta. \tag{10}$$

As regards the hadronic tensor, the generalized factorization theorem[34, 23, 35] in the covariant formalism[36] yields, at zero order in the QCD coupling constant,

$$H_{\mu\nu} = \frac{1}{3}\sum_{f=1}^{6} e_f^2 \int d\Gamma_q \varphi^f(p'; P) h_{\mu\nu}^f(p, p'; S), \tag{11}$$

$$h_{\mu\nu}^f = \sum_L q_L^f(p) Tr(\rho^L \gamma_\mu \rho' \gamma_\nu). \tag{12}$$

Here the factor $1/3$ comes from color averaging in the elementary scattering process and $f$ runs over the three light flavors $(u, d, s)$ and antiflavors $(\overline{u}, \overline{d}, \overline{s})$, $e_1 = -e_4 = 2/3$, $e_2 = e_3 = -e_5 = -e_6 = -1/3$. $p$ and $p'$ are respectively the four-momenta of the active parton before and after being struck by the virtual photon. $S$ is the Pauli-Lubanski (PL) four-vector of the proton. $q_L^f(p)$ is the probability density of finding a quark (or an antiquark) with four-momentum $p$, whose spin is parallel ($L = 1/2$) or antiparallel ($L = -1/2$) to the proton spin. Analogously $\varphi^f(p')$ is the fragmentation function of a quark of four-momentum $p'$ into a pion of four-momentum $P$. Moreover

$$d\Gamma_q = \frac{1}{(2\pi)^2} d^4p \, \delta(p^2)\theta(p_0) d^4p' \, \delta(p'^2)\theta(p'_0) \, \delta^4(p'-p-q), \tag{13}$$

the active parton being taken on shell and massless. Lastly the $\rho$'s are the spin density matrices of the initial and final active parton, $i.\ e.$[29],

$$\rho^L = \frac{1}{2}\slashed{p}[1 + 2L\gamma_5(\lambda + \slashed{\eta})] \quad \text{and} \quad \rho' = \frac{1}{2}\slashed{p}'. \tag{14}$$

Here $2L\eta$ is the transverse PL four-vector of the active parton, while $\lambda$ is the longitudinal component of the quark spin vector. Formulae (14) are consistent with the Politzer theorem[37] in the parton model approximation. These imply, together with eq. (12), that $\eta$ does not contribute to $h^f_{\mu\nu}$. For later convenience we re-write this last tensor as

$$h^f_{\mu\nu} = \frac{1}{4}\left[q^f(p)s_{\mu\nu} + \lambda \delta q^f(p) a_{\mu\nu}\right]. \tag{15}$$

Here

$$s_{\mu\nu} = Tr(\slashed{p}\gamma_\mu \slashed{p}'\gamma_\nu), \qquad a_{\mu\nu} = Tr(\gamma_5 \slashed{p}\gamma_\mu \slashed{p}'\gamma_\nu). \tag{16}$$

Moreover $q^f(p) = \sum_{L=\pm 1/2} q^f_L(p)$ is the unpolarized quark distribution and $\delta q^f(p) = \sum_{L=\pm 1/2} 2L q^f_L(p)$. $\lambda$ is a Lorentz invariant, pseudoscalar quantity, such that $|\lambda| \leq 1$. If we neglect the parton transverse momentum, the only way of constructing such a quantity with the available vectors is

$$\lambda = \lambda_\parallel = \mathbf{S} \cdot \frac{\mathbf{q}}{|\mathbf{q}|} = \frac{-S \cdot q}{\sqrt{\nu^2 + Q^2}}. \tag{17}$$

Here we have exploited the fact that $\nu$ is a Lorentz scalar and that in the laboratory frame $q \equiv (\nu, \mathbf{q})$ and $S \equiv (0, \mathbf{S})$, where $\mathbf{S}$ is the proton spin vector, $\mathbf{S}^2 = 1$. $\lambda_\parallel$ can be viewed as the helicity of the proton in a frame where the proton is moving along $\mathbf{q}$. In this connection we observe that the spatial direction of the virtual photon can also be defined covariantly by means of the four-momenta of the photon and of the proton[19]. In order to take into account the transverse momentum, we have to adopt a frame where the proton momentum is large in comparison to $M$[38]. To this end we consider the Breit frame - coincident with the one adopted by Feynman[39] - where the virtual photon has a four-momentum $q = (0, \mathbf{q}_B)$, with $|\mathbf{q}_B| = Q$. In this frame the proton momentum is $-\frac{1}{2x}\mathbf{q}_B$, therefore the active parton carries a momentum $\mathbf{p}_B = -\frac{1}{2}\mathbf{q}_B + \mathbf{p}_\perp$, where, as usual, $x = Q^2/2M\nu$ is the longitudinal fractional momentum and $\mathbf{p}_\perp$ the transverse momentum with respect to $\mathbf{q}_B$. We decompose the proton spin vector $\mathbf{S}$ into a longitudinal and a transverse component, i. e.,

$$\mathbf{S} = \lambda_\parallel \frac{\mathbf{q}}{|\mathbf{q}|} + \mathbf{S}_\perp, \qquad \mathbf{S}_\perp \cdot \mathbf{q} = 0. \tag{18}$$

But the average spin of the quark is independent of the quantization axis, therefore the decomposition (18) implies

$$\lambda \delta q^f(x, \mathbf{p}_\perp) = \lambda_\parallel \delta q^f_\parallel(x, \mathbf{p}_\perp^2) + \lambda_\perp \delta q^f_\perp(x, \mathbf{p}_\perp). \tag{19}$$

Here

$$\lambda_\perp = \frac{\mathbf{S}_\perp \cdot \mathbf{p}_B}{|\mathbf{p}_B|} \simeq \frac{2\mathbf{S} \cdot \mathbf{p}_\perp}{Q}, \qquad (20)$$

moreover $\delta q_\parallel^f(x, \mathbf{p}_\perp^2)$ (denoted as $g_{1L}(x, \mathbf{p}_\perp^2)$ by MT) is the t.m.d. helicity distribution. It is interesting to write the relationships of $\delta q_\parallel^f$ and $\delta q_\perp^f$ to the usual distributions:

$$\Delta q^f(x) = \int d^2 p_\perp \delta q_\parallel^f(x, \mathbf{p}_\perp^2); \qquad h_1^f(x) = \int d^2 p_\perp \delta q_\perp^f(x, \mathbf{p}_\perp). \qquad (21)$$

Now we carry on the integration (11) over the time and longitudinal components of $p$, taking the $z$-axis opposite to $\mathbf{q}$. We get, in the light cone formalism,

$$H_{\mu\nu} = \frac{1}{4\pi^2 Q^2} \sum_{f=1}^{6} e_f^2 \int d^2 p_\perp \varphi^f(z, \mathbf{P}_\perp^2) h_{\mu\nu}^f(x, \mathbf{p}_\perp; \mathbf{S}), \qquad (22)$$

where the tensor $h_{\mu\nu}^f$ reads, after insertion of eq. (19) into eq. (15),

$$h_{\mu\nu}^f = \frac{1}{4} \left\{ q^f(x, \mathbf{p}_\perp^2) s_{\mu\nu} + \left[ \lambda_\parallel \delta q_\parallel^f(x, \mathbf{p}_\perp^2) + \lambda_\perp \delta q_\perp^f(x, \mathbf{p}_\perp) \right] a_{\mu\nu} \right\}. \qquad (23)$$

$q^f$ and $\varphi^f$ are related, respectively, to the usual unpolarized distribution and to the pion fragmentation function, i. e.,

$$q^f(x) = \int d^2 p_\perp q^f(x, \mathbf{p}_\perp^2), \qquad D^f(z) = \int d^2 P_\perp \varphi^f(z, \mathbf{P}_\perp^2). \qquad (24)$$

Moreover $z = (P_\parallel + P_0)/(2|\mathbf{p}'|)$ is the longitudinal fractional momentum of the pion resulting from fragmentation of the struck parton, whose momentum is $\mathbf{p}'$. We have defined $P_0 = \sqrt{m_\pi^2 + \mathbf{P}^2}$, $P_\parallel = \mathbf{P} \cdot \mathbf{p}'/|\mathbf{p}'|$ and $\mathbf{P}_\perp = \mathbf{P} - P_\parallel \mathbf{p}'/|\mathbf{p}'|$, $\mathbf{P}$ being the pion momentum in the laboratory frame. Denoting by $\mathbf{\Pi}_\perp$ the transverse momentum of the pion with respect to the photon momentum, we get

$$\mathbf{P}_\perp = \mathbf{\Pi}_\perp - z\mathbf{p}_\perp. \qquad (25)$$

Therefore, if we keep $\mathbf{\Pi}_\perp$ fixed, $\mathbf{P}_\perp$ depends on $\mathbf{p}_\perp$.

We notice that, although we have chosen a particular frame, the result (22) is apparently covariant.

## 2.2 Azimuthal asymmetry

In order to calculate the double spin azimuthal asymmetry $A(|\mathbf{k}|; Q, \nu; \mathbf{P})$ for reaction (4) - defined analogously to (5), but keeping the pion momentum $\mathbf{P}$ fixed - we have to substitute the leptonic tensor (10) and the hadronic tensor (22) into the cross section (7), taking into account relations (23) and (20). The result is

$$A(|\mathbf{k}|; Q, \nu; \mathbf{P}) = \mathcal{F} \frac{\sum_{f=1}^{6} e_f^2 \delta Q^f}{\sum_{f=1}^{6} e_f^2 Q^f}, \qquad \mathcal{F} = \frac{k_+ k'_- - k_- k'_+}{k_+ k'_- + k_- k'_+}. \qquad (26)$$

$\mathcal{F}$ is the depolarization of the virtual photon with respect to the parent lepton[21]. Moreover we have introduced the quantities

$$Q^f = Q^f(x, z, \mathbf{\Pi}_\perp^2) = \int d^2 p_\perp q^f(x, \mathbf{p}_\perp^2) \varphi^f(z, \mathbf{P}_\perp^2), \qquad (27)$$

$$\delta Q^f = \delta Q^f_\parallel(x, z, \mathbf{\Pi}_\perp^2) + \mathbf{\Pi}_\perp \cdot \mathbf{S} \delta Q^f_\perp(x, z, \mathbf{\Pi}_\perp), \qquad (28)$$

$$\delta Q^f_\parallel(x, z, \mathbf{\Pi}_\perp^2) = \lambda_\parallel \int d^2 p_\perp \delta q^f_\parallel(x, \mathbf{p}_\perp^2) \varphi^f(z, \mathbf{P}_\perp^2), \qquad (29)$$

$$\delta Q^f_\perp(x, z, \mathbf{\Pi}_\perp) \mathbf{\Pi}_\perp \cdot \mathbf{S} = \int d^2 p_\perp \lambda_\perp \delta q^f_\perp(x, \mathbf{p}_\perp) \varphi^f(z, \mathbf{P}_\perp^2). \qquad (30)$$

The pseudoscalar character of $\delta Q^f$ (eq. (28)) follows from assuming a massless lepton: indeed, the expression we have deduced for the asymmetry (see the first eq. (26)) holds in any frame where the lepton mass is negligible and the lepton helicity has the same value as in the laboratory frame. Below we shall show that $\delta Q^f$ is twist 3. From formula (28) we deduce that, in order to maximize the contribution of $\delta q^f_\perp$ to our asymmetry, one has to take the vector $\mathbf{\Pi}_\perp$ parallel to or opposite to $\mathbf{S}$, that is, to select pions whose momenta lie in the $(\mathbf{q}, \mathbf{S})$ plane. Furthermore eq. (28) implies that $\delta Q^f$ is especially sensitive to $\delta q^f_\perp(x, \mathbf{p}_\perp)$ if $\mathbf{q} \cdot \mathbf{S} = 0$. Indeed, in this situation the first term of eq. (28) - and more generally the JJ asymmetry - vanishes. Therefore events such that the lepton scattering plane is perpendicular to the proton polarization are particularly relevant to our aims.

Since the products $k_+ k'_-$ and $k_- k'_+$ are invariant under boosts along the $z$-axis, we calculate them in the laboratory frame, where

$$k_\pm = \frac{|\mathbf{k}|}{\sqrt{2}}(1 \pm cos\beta), \qquad k'_\pm = \frac{|\mathbf{k}'|}{\sqrt{2}}[1 \pm cos(\theta + \beta)]. \qquad (31)$$

Here $\mathbf{k}' = \mathbf{k} - \mathbf{q}$ is the final lepton momentum. Moreover $\beta$ and $\theta$ are, respectively, the angle between $\mathbf{k}$ and $\mathbf{q}$ and between $\mathbf{k}$ and $\mathbf{k}'$:

$$|\mathbf{q}|cos\beta = |\mathbf{k}| - |\mathbf{k}'|cos\theta. \tag{32}$$

Now we consider the scaling limit, i. e., $Q^2 \to \infty$, $\nu \to \infty$, $Q^2/2M\nu \to x$. Since $Q^2 \simeq 2|\mathbf{k}||\mathbf{k}'|(1 - cos\theta)$, $\theta$ tends to zero in that limit, as well as $\beta$:

$$\theta \simeq \frac{M}{Q}\frac{y}{x(1-y)^{1/2}}, \qquad \beta \simeq \theta\frac{1-y}{y}, \qquad y = \frac{\nu}{|\mathbf{k}|}. \tag{33}$$

Then the second eq. (26) and eq. (17) yield, respectively,

$$\mathcal{F} = \frac{y(2-y)}{1+(1-y)^2}, \qquad \lambda_\parallel = \frac{1-y}{y}sin\theta cos\phi', \tag{34}$$

where $\phi'$ is the azimuthal angle between the $(\mathbf{k}, \mathbf{k}')$ plane and the $(\mathbf{k}, \mathbf{S})$ plane. Therefore $\delta Q_\parallel^f$, eq. (29), is twist 3, as follows from the second eq. (34) and from the first eq. (33). But also the second term of eq. (28) is twist 3, as is immediate to check. Therefore our asymmetry is twist 3.

## 2.3 Numerical estimates

Now we calculate the order of magnitude of the asymmetry (26) under optimal conditions. To this end, first of all, according to the considerations of subsection 2.2, we take into account events such that the azimuthal angle $\phi'$ (see the second eq. (34)) is about $\pi/2$ and $\mathbf{\Pi}_\perp$ is parallel (or antiparallel) to $\mathbf{S}$. Moreover, eqs. (28) to (30) and eqs. (34) suggest that $y$ should be chosen as close as possible to 1. Assuming the transverse momentum distributions involved in asymmetry (26) to be of the Gaussian type[†], with equal widths of $\sim 0.85\ GeV/c$, and setting $|\mathbf{\Pi}_\perp| \simeq 1\ GeV$ and $Q = 2.5\ GeV$, the asymmetry (26) results in $A \sim 0.4R$, where $R = h_1^f(x)/q^f(x)$ has been determined by HERMES[11], $|R| = (50 \pm 30)\%$.

## 2.4 Remarks

At this point some remarks are in order.

---

[†]see eqs. (37) to (39) below

(i) Parity invariance and a rotation of the reference frame by $\pi$ around the proton spin imply

$$\delta q_\perp^f(x, \mathbf{p}_\perp) = \delta q_\perp^f(x, -\mathbf{p}_\perp), \qquad (35)$$

the typical property of a T-even function[40]. This, in turn, implies that, if we integrate the SIDIS differential cross section - and therefore the hadronic tensor (22) - over $\mathbf{\Pi}_\perp$, the second term of eq. (28) vanishes. On the other hand, upon integration, the first term of eq. (28) goes over into $\lambda_\| \Delta q^f(x) D^f(z)$, corresponding to the "kinematic" twist-3 term of the numerator in the JJ asymmetry[27]. The above mentioned numerator includes also a "dynamic" twist-3 term of the type $\lambda_\| [\frac{1}{x} h_1^f(x) \frac{1}{z} \hat{e}^f(z) + g_T^f(x) D^f(z)][27]$ (see introduction), where $g_T^f(x)$ is the transverse spin distribution. This term does not appear in asymmetry (26), since we limit ourselves to the contributions of the QCD parton model.

(ii) Eqs. (26) to (30) and the first eq. (34) hold true for any orientation of the proton spin. However, if $\mathbf{S}$ is not perpendicular to the lepton beam, $\lambda_\|$ does not decrease with $Q$. Therefore if, e. g., the proton is polarized longitudinally, the asymmetry derives still a contribution from the t.m.d. tranversity function, which, however, risks to be taken over by the JJ term[22]. We shall see in the next section a method for extracting $\delta q_\perp^f$ under such unfavourable conditions, as, e. g., in the HERMES experiment[11].

(iii) The hadronic tensor (22), which we have derived starting from the definition of transversity by JJ1, turns out to coincide with the tensor found by KM, provided we assume[‡]

$$g_{1T}^f = \frac{2M}{Q} \delta q_\perp^f. \qquad (36)$$

This relationship is proven in appendix, together with other useful connections among the t.m.d. distributions defined by MT (see also ref.[41]). Therefore KM's considerations and deductions about $g_{1T}^f$ - like, e. g., its relation with $g_2^f$ - can be applied to $\delta q_\perp^f$.

---

[‡]according to the normalization constant chosen by KM for $g_{1T}^f$

# 3 Extracting transversity from data

In this section we discuss how to extract $\delta q_\perp^f$ from data. To this end we illustrate two different methods: a best fit to the asymmetry, with a suitable parametrization for $\delta q_\perp^f$, and the use of weighted asymmetries[21, 19]. Furthermore we suggest how to take into account the spurious effects, that is, contributions to asymmetry (26) not related to $\delta q_\perp^f$.

## 3.1 Gaussian parametrization

A frequently used parametrization of the t.m.d. unpolarized distributions and fragmentation functions consists of[16, 23, 35]

$$q^f(x, \mathbf{p}_\perp^2) = (a/\pi) q^f(x) exp(-a\mathbf{p}_\perp^2), \tag{37}$$

$$\varphi^f(z, \mathbf{P}_\perp^2) = (a_\pi/\pi) D^f(z) exp(-a_\pi \mathbf{P}_\perp^2). \tag{38}$$

Here $a \sim 1.38~(GeV/c)^{-2}$, as results from DY[42], and $a_\pi$ may be determined from two-jet events in $e^+e^- \to \pi X$. As regards the t.m.d. transversity, it looks appropriate to set[43]

$$\delta q_\perp^f(x, \mathbf{p}_\perp) = (a/\pi) h_1^f(x) exp(-a\mathbf{p}_\perp^2). \tag{39}$$

Here $h_1^f$ may be parametrized according to the suggestion of ref.[44], i. e.,

$$h_1^f(x) = \frac{1}{2} N \frac{x^\alpha (1-x)^\beta}{\alpha^\alpha \beta^\beta} (\alpha+\beta)^{\alpha+\beta} \left[ q^f(x) + \Delta q^f(x) \right], \tag{40}$$

where $N$, $\alpha$, $\beta$ and $b$ are free parameters, with $|N| \leq 1$.

## 3.2 Weighted asymmetries

### 3.2.1 The KM weight function

For the reaction we are studying, KM have defined a weighted asymmetry of the type (3), suggesting the weight function $w(\mathbf{\Pi}_\perp) = 2\mathbf{\Pi}_\perp \cdot \mathbf{S}/M$. Inserting this function and eq. (7) into eq. (3), and taking into account eqs. (10) and (22), we get

$$\langle A_w \rangle = z \mathcal{F} \frac{\sum_{f=1}^{6} e_f^2 h_1^{f(1)}(x) D^f(z)}{\sum_{f=1}^{6} e_f^2 q^f(x) D^f(z)}, \tag{41}$$

where
$$h_1^{f(1)}(x) = \frac{2}{QM} \int d^2p_\perp (\mathbf{p}_\perp \cdot \mathbf{S})^2 \delta q_\perp^f. \tag{42}$$

The expression KM obtain for the weighted asymmetry differs from eq. (41) by the substitution $h_1^{f(1)}(x)$ by $g_{1T}^{f(1)}(x)$, where

$$g_{1T}^{f(1)}(x) = \frac{1}{2M^2} \int d^2p_\perp \mathbf{p}_\perp^2 g_{1T}^f. \tag{43}$$

But eq. (36) implies $g_{1T}^{f(1)}(x) = h_1^{f(1)}(x)$, so that the KM asymmetry turns out to coincide with eq. (41). Moreover the numerical evaluation of $g_{1T}$ that KM infer from E143 data[45] allows, by means of eqs. (42) and (39), to evaluate $h_1$. This behaves quite similarly to the bag model prediction (see, e. g., JJ1), especially at small $x$. The accord between the two calculations can be made quantitative for $Q$ of order 12 to 15 $GeV$.

### 3.2.2 Harmonic oscillator weight functions

More refined information on the t.m.d. transversity function could be extracted from data by using a different set of weight functions, inspired to the Gaussian parametrization. We assume for $\varphi^f$ and $\delta q_\perp^f$ expansions of the type

$$\varphi^f(z, \mathbf{P}_\perp^2) = \sum_K c_K^f(z) \Phi_K(P), \qquad \delta q_\perp^f(x, \mathbf{p}_\perp) = \sum_K \tilde{c}_K^f(x) \tilde{\Phi}_K(p). \tag{44}$$

Here $P = (\mathbf{P}_\perp^2)^{1/2}$, $p = (\mathbf{p}_\perp^2)^{1/2}$ and the $\Phi_K(P)$ are the eigenfunctions of the equation

$$\left[ a_\pi^2 P^2 + \frac{1}{4}\xi^2 - a_\pi(K + \frac{1}{2}) \right] \Phi_K(P) = 0, \tag{45}$$

where $\xi = id/dP$. The $\tilde{\Phi}_K$ fulfil an analogous equation, with $a$ instead of $a_\pi$. These functions have been chosen in such a way that the terms with $K = 0$ reproduce the parametrizations of subsection 3.1. $c_K^f(z)$ and $\tilde{c}_K^f(x)$ are real coefficients. The set of weight functions we propose is

$$w_K = 2 \frac{\mathbf{\Pi}_\perp \cdot \mathbf{S}}{M} \Phi_K(|\mathbf{\Pi}_\perp|). \tag{46}$$

We denote by $\langle A_w^K \rangle$ the weighted asymmetries obtained by substituting the functions (46) into eq. (3). Taking into account eqs. (44) and once more eqs. (7), (10) and

(22), we get

$$\langle A_w^K \rangle \sum_{f=1}^{6} e_f^2 q^f(x) D^f(z) = \mathcal{F} \sum_{f=1}^{6} \sum_L e_f^2 M_{KL}^f(z) \tilde{c}_L^f(x), \qquad (47)$$

where

$$M_{KL}^f(z) = \frac{4}{QM} \int d^2 P_\perp \mathbf{P}_\perp \cdot \mathbf{S} \Phi_K(P) \int d^2 p_\perp \mathbf{p}_\perp \cdot \mathbf{S} \tilde{\Phi}_L(p) \varphi^f \left[ z, (\mathbf{P}_\perp - z\mathbf{p}_\perp)^2 \right]. \qquad (48)$$

The linear system (47) can be solved with respect to $\tilde{c}_L^f(x)$, provided we assume the dominance of some fragmentation mechanism[27], which reduces the sum over $f$ to a single term. Alternatively, if data relative to asymmetry for $\pi^\pm$ and to $K$-mesons are simultaneously available, the system (47) can be written as

$$\langle A_w^{KF} \rangle \sum_{f=1}^{6} e_f^2 q^f(x) D_F^f(z) = \mathcal{F} \sum_{f=1}^{6} \sum_L e_f^2 M_{KL}^{fF}(z) \tilde{c}_L^f(x), \qquad (49)$$

where $F$ runs over the final hadrons $\pi^+$, $\pi^-$ and $K$. This new system can be solved if we make some assumptions, so as to reduce the sum over $f$ to three or less terms. For example, we may consider separately small and large $x$. In the latter $x$-interval we may neglect sea contribution and the system is overdetermined, since $f$ runs over two flavors. For small $x$, where the sea prevails, we may solve the system by assuming a relation between quark and antiquark distributions, in such a way that $f$ run over three flavors. Once the coefficients $\tilde{c}_L^f(x)$ have been determined for any $f$ and for a sufficiently large number of $L$, we may determine $h_1^f$ thanks to the second eq. (44) and to the second eq. (21). The method we have just suggested is somewhat similar to the purity method used by HERMES in splitting the single flavor contribution[46] in longitudinally polarized distributions (see also refs.[31, 32, 47, 48, 49, 50]).

The weighted cross sections $\langle A_w^K \rangle$ wash out the unwanted JJ contribution, therefore they are particularly suitable in an experiment, like HERMES[11], where the target is longitudinally polarized.

## 3.3 Spurious effects

Asymmetry (26), as well as any asymmetry connected to a SIDIS experiment, is sensitive to processes which have nothing to do with the distribution we want to determine. Such processes, whose contributions may be perhaps reduced, but not eliminated by means of kinematic cuts[31], are essentially of two kinds:

(i) interactions[30] and/or correlations[32] between the active quark and the spectator partons;

(ii) fragmentation of the target remnant diquark or decay of a light cluster[31].

Gluon exchange between the active quark and the spectator partons is demanded by gauge invariance[34]. A gluon emitted by spectator partons may interact with the active quark *a)* before or *b)* after being struck by the photon. The amplitude for such a process interferes with the amplitude for the active parton to interact with the sole photon, giving rise to a "dynamic" twist-3 contribution[28, 29] (see also refs. [30, 51]). In our case this term must be T-even, since T-odd functions[30, 51] are involved only in single spin asymmetries. In particular, the type-*a)* contribution corresponds to $g_T^f(x, \mathbf{p}_\perp)$, whose integral over the transverse momentum is $g_T^f(x)$. Calculations[29] within the model proposed by Qiu and Sterman[34] assure that such contributions are about 10% of the parton model term.

The spurious fragmentation effects (ii)[31] and quark-spectator correlations[32] are nonperturbative and can be taken into account according to a strategy similar to the one followed by Kotzinian[31]. The method consists of a simulation program based on the Lund fragmentation model. The simulation[52, 53] reproduces the data of unpolarized SIDIS, but exhibits deviations from the purity assumption[46, 47, 48, 49, 50] that the final hadron observed come exclusively from the active quark fragmentation.

Such deviations are suitably described by fracture functions[26]. In our case the asymmetry (26) has to be modified into

$$A'(|\mathbf{k}|; Q, \nu; \mathbf{P}) = \mathcal{F} \; \frac{\sum_{f=1}^{6} \left[ e_f^2 \delta Q^f + \delta \mathcal{M}^f \right]}{\sum_{f=1}^{6} \left[ e_f^2 Q^f + \mathcal{M}^f \right]}. \tag{50}$$

Here $\mathcal{M}^f$ and $\delta \mathcal{M}^f$ are respectively the unpolarized and polarized fracture functions, which depend on $x$, $z$ and $\mathbf{\Pi}_\perp$. As in the case of longitudinal polarization[31], the cross section for reaction (4) can be calculated by means of the simulation program, assuming parametrizations (37) to (39) for the nonperturbative functions involved, and using some model calculation for $h_1^f$[2, 20, 44]. Since the program is endowed with a pointer, which shows the origin of the final hadrons[31], one can determine the fracture functions by selecting the spurious events that result from the simulation, and fitting them by means of a suitable parametrization, e. g., $\delta\mathcal{M}^f = C\delta Q^f$, $C$ being a constant.

Incidentally, in the case of longitudinal polarization, these spurious effects might be responsible for the discrepancy between the recent HERMES results[46] and expectations from the so-called "spin crisis"[31].

It is worth noticing that the weighted asymmetries wash out in part such effects, for example the light cluster decays, which are isotropic.

# 4 Comparison with other methods

Extraction of $h_1$ from SIDIS single spin asymmetry is affected by drawbacks similar to those described in subsect. 3.3. In particular this asymmetry derives a contribution from the Sivers effect[33], which could be even greater than the term we are interested in. In fact the Sivers function is chiral even and therefore is coupled to the usual t.m.d. fragmentation function; this is considerably larger than $\kappa$, the chiral odd and T-odd fragmentation function coupled to $h_{1T}$ in the single spin asymmetry (see sect. 1). Moreover there are suggestions that this single spin asymmetry, usually classified as twist 2[25], could be instead twist 3[54, 26], as well as asymmetry (26). Indeed, this latter is twist 3 (see subsect. 2.2), while it was considered twist 2 in previous papers[16, 23, 21]. The reason for this discrepancy is explained in appendix. There we discuss a factor $\mu^{-1}$, introduced for dimensional reasons, where $\mu$ is an undetermined energy scale. This is usually assumed equal to $M$[1, 16], whereas a correct normalization of the t.m.d. distributions as probability densities demands $\mu = Q/2$. Lastly, in our method, the complication of inferring $h_1$ from a number of weighted asymmetries, as described in subsect. 3.2, is largely compensated by the fact that $g_{1T}$ is coupled to the ordinary fragmentation function of the pion, whose parameters are well determined, contrary to the Collins function.

Among the other methods proposed for extracting transversity, we recall proton-proton collisions, with muon pairs or hadrons (pions or lambdas) in the final state[55], and SIDIS with a lambda in the final state[14, 15]. But asymmetries in proton-proton collisions have been shown to be rather insensitive to $h_1$[55]. Concerning SIDIS, string fragmentation into a lambda is much more difficult than pion fragmentation; this not only makes statistics poorer, but also increases the drawbacks described in subsect. 3.3[26].

# 5 Conclusion

To conclude, $\delta q_\perp$ may be coupled to a chiral-even fragmentation function, in particular to the twist-2, unpolarized t.m.d. fragmentation function of the final hadron observed. This suggests an alternative, convenient method for extracting $h_1$, circumventing some of the usual drawbacks that plague determination of transversity. Specifically, we propose to measure the double spin azimuthal asymmetry in semi-inclusive pion leptoproduction. For reasonable values of $Q^2$ (4 to 10 $GeV^2$), and under the most favourable kinematic conditions, the asymmetry is estimated to be at least $\sim 10\%$. The suggested experiment could be performed at facilities like CERN (COMPASS coll.) and DESY (HERMES coll.), where similar asymmetry measurements are being realized or planned. As a last comment, our results confirm the crucial role of the intrinsic transverse momentum[16, 17, 38, 40] and of polarized SIDIS experiments[47, 48, 49, 50] in extracting quark distributions.

### Acknowledgments

The author is thankful to his friends M. Anselmino, A. Di Giacomo, E. Leader, P. Mulders and O. Teryaev for useful and stimulating discussions.

## Appendix

Here we establish some relationships among transverse momentum dependent (t.m.d.) distributions, for which we adopt the notations by Mulders and Tangerman[16] (MT) (see also Ralston and Soper[1] (RS) and other authors[17, 21]). In particular we show that, in the chiral limit, $g_{1T}$ equals $h_{1T}$ ($\delta q_\perp$ in the present paper). First of all, we write the correlation matrix in the QCD parton model. Secondly we compare it with the general expression of the correlation matrix. As a result we find some relationships among t.m.d. distribution functions, which turn out to hold true even taking into account quark-gluon interactions and renormalization effects. We give also an alternative proof of the relationship between $g_{1T}$ and $h_{1T}$ in the chiral limit. Lastly, we shortly discuss our result.

### A.1 - Correlation matrix in QCD parton model

*A.1.1 - The correlation matrix*

We define the correlation matrix as

$$\Phi(x, p_\perp; P, S) = \int dp^- \int \frac{d^4y}{(2\pi)^4} e^{ipy} \langle P, S | \psi(y) \overline{\psi}(0) | P, S \rangle, \qquad (A.1)$$

assuming the light cone gauge $A^+ = 0$. Here $P$ and $S$ are, respectively, the four-momentum and the Pauli-Lubanski (PL) four-vector of the proton, with $S^2 = -1$; $\psi$ is the quark field. $p$ is the quark four-momentum, whose transverse component with respect to the proton momentum is $p_\perp$, i. e.,

$$\begin{aligned} p &= p^+ n_+ + p_\perp + p^- n_-, & P &= P^+ n_+ + P^- n_-, & \text{(A.2)} \\ n_+^2 &= n_-^2 = 0, & n_+ \cdot n_- &= 1, & \text{(A.3)} \\ n_+ \cdot p_\perp &= n_- \cdot p_\perp = 0. & & & \text{(A.4)} \end{aligned}$$

Moreover $x = p^+/P^+$ is the light cone fraction of the quark momentum.

We take a frame such that the proton has a large momentum, say $\mathcal{P}$, much greater than its rest mass $M$, and it is transversely polarized. Moreover we choose the $z$-axis along the proton momentum and the $y$-axis along the proton polarization, so that

$$\begin{aligned} P &\equiv (\sqrt{M^2 + \mathcal{P}^2}, 0, 0, \mathcal{P}), & S &\equiv (0, 0, 1, 0), & \text{(A.5)} \\ n_\pm &\equiv \frac{1}{\sqrt{2}}(1, 0, 0, \pm 1), & p_\perp &\equiv (0, \mathbf{p}_\perp), & \text{(A.6)} \\ \mathbf{p}_\perp &\equiv (p_1, p_2, 0), & |\mathbf{p}_\perp| &= O(M). & \text{(A.7)} \end{aligned}$$

From now on this frame will be called $\mathcal{P}$-frame. For the sake of simplicity, we exclude T-odd terms in the correlation matrix (A.1).

*A.1.2 - QCD-improved parton model*

The correlation matrix for free, on-shell[23] quarks in a transversely polarized proton reads

$$\Phi_\perp^{free} = \sum_{T=\pm 1/2} q_T(x, \mathbf{p}_\perp)\frac{1}{2}(\not{p} + m)(1 + 2T\gamma_5 \not{S}_q). \tag{A.8}$$

Here $m$ and $p$ are, respectively, the rest mass and the four-momentum of the quark, such that $p^2 = m^2$. $2TS_q$ is the *quark* PL vector, with $S_q^2 = -1$. $q_T(x, \mathbf{p}_\perp)$ is the probability density of finding a quark with its spin aligned with ($T = 1/2$) or opposite to ($T = -1/2$) the proton spin. Now we define a *quark* rest frame[38] - to be named $q$-frame from now on - whose axes are parallel to those of the $\mathcal{P}$-frame. In this frame we have $S_q = S_q^{(0)} = S \equiv (0, 0, 1, 0)$. Decomposing $S_q^{(0)} = S$ into a transverse and a longitudinal component with respect to the quark momentum, we get

$$S_q^{(0)} = S = \Sigma_\perp cos\theta' + \nu sin\theta'. \tag{A.9}$$

Here

$$sin\theta' = sin\theta sin\phi, \qquad sin\theta = \frac{|\mathbf{p}_\perp|}{|\mathbf{p}|}, \qquad sin\phi = \frac{-p_\perp \cdot S}{|\mathbf{p}_\perp|}, \tag{A.10}$$

$$\mathbf{p} \simeq (\mathbf{p}_\perp, x\mathcal{P}), \qquad \nu \equiv (0, \mathbf{t}), \qquad \Sigma_\perp \equiv (0, \mathbf{n}), \tag{A.11}$$

$$\mathbf{t} \equiv (sin\theta cos\phi, sin\theta sin\phi, cos\theta), \tag{A.12}$$

$$\mathbf{n} \equiv (cos\theta cos\phi, cos\theta sin\phi, -sin\theta). \tag{A.13}$$

In order to calculate $S_q$ in the $\mathcal{P}$-frame, we perform a boost along the quark momentum. This boost leaves $\Sigma_\perp$ invariant and transforms $\nu$ into $\tilde{p}/m$, where

$$\tilde{p} \equiv (|\mathbf{p}|, E_q \mathbf{t}), \qquad E_q = \sqrt{m^2 + \mathbf{p}^2}. \tag{A.14}$$

As a result we get

$$S_q = S + \left[\frac{p}{m} - (\delta + \nu)\right] sin\theta'. \tag{A.15}$$

Here we have defined

$$\delta = \frac{m}{\sqrt{2}x\mathcal{P}}n'_{-}\left[1+O(\mathcal{P}^{-2})\right], \qquad n'_{\pm} \equiv \frac{1}{\sqrt{2}}(1, \pm\mathbf{t}). \qquad (A.16)$$

We stress that it is essential that the relative momentum of the quark with respect to the proton is nonzero; otherwise $x$ would be fixed and equal to $m/M$[38]. Substituting eq. (A.15) into eq. (A.8), and taking into account the definitions (A.11) and (A.16) of $\nu$ and $\delta$, we get

$$\begin{aligned}\Phi_{\perp}^{free} &= \frac{1}{2}q(x, \mathbf{p}_{\perp}^2)(\slashed{p}+m) \\ &+ \frac{1}{2}\delta q_{\perp}(x, \mathbf{p}_{\perp})\gamma_5\left\{\frac{1}{2}[\slashed{S}, \slashed{p}] + \slashed{p}\sin\theta' - A + mB\right\} + O(\mathcal{P}^{-1}).\end{aligned} \qquad (A.17)$$

Here we have set

$$q(x, \mathbf{p}_{\perp}^2) = \sum_{T=\pm 1/2} q_T(x, \mathbf{p}_{\perp}) \qquad \delta q_{\perp}(x, \mathbf{p}_{\perp}) = \sum_{T=\pm 1/2} 2T q_T(x, \mathbf{p}_{\perp}), \qquad (A.18)$$

$$A = E_q \frac{1}{2}[\slashed{n}'_+, \slashed{n}'_-]\sin\theta', \qquad (A.19)$$

$$B = \slashed{S} + \frac{1}{\sqrt{2}}\left\{\slashed{n}'_-\left(1-\frac{m}{|\mathbf{p}|}\right) - \slashed{n}'_+ + \frac{1}{\sqrt{2}}[\slashed{n}'_+, \slashed{n}'_-]\right\}\sin\theta'. \qquad (A.20)$$

Moreover we have exploited the relations $-\slashed{p}\slashed{S} = 1/2[\slashed{S}, \slashed{p}] - p\cdot S$, $p\cdot S = p_{\perp}\cdot S$ and $\nu = \frac{1}{\sqrt{2}}(n'_+ - n'_-)$.

### A.2 - Parametrization of the correlation matrix

Now we parametrize the correlation matrix of a transversely polarized proton in the $\mathcal{P}$-frame. We normalize this matrix in such a way that the leading twist contribution of $\Phi$ for an unpolarized proton, when integrated over $\mathbf{p}_{\perp}$, coincide with the product of the usual density matrix times the distribution of an unpolarized quark. Taking into account the operators of the Dirac algebra[23, 16, 24], we get

$$\Phi_{\perp} = \Phi_{0a} + \Phi_{0b} + \Phi_1 + \Phi_2. \qquad (A.21)$$

Here

$$\Phi_{0a} = \frac{1}{\sqrt{2}}x\mathcal{P}\left(f_1\slashed{n}_+ + \lambda_{\perp}g_{1T}\gamma_5\slashed{n}_+ + \frac{1}{2}h_{1T}\gamma_5[\slashed{S}, \slashed{n}_+]\right)$$

$$+ \frac{1}{4\sqrt{2}} \lambda_\perp h_{1T}^\perp \gamma_5 [\slashed{p}_\perp, \slashed{n}_+], \tag{A.22}$$

$$\Phi_{0b} = \frac{1}{2}\left(f_1^\perp + \lambda_\perp g_T^\perp \gamma_5\right) \slashed{p}_\perp$$

$$+ \frac{1}{4}\lambda_\perp \left(h_T^\perp \gamma_5 [\slashed{S}, \slashed{p}_\perp] + h_T \mu \gamma_5 [\slashed{n}_-, \slashed{n}_+]\right), \tag{A.23}$$

$$\Phi_1 = \frac{1}{2} M \left(e + g_T \gamma_5 \slashed{S}\right), \tag{A.24}$$

while $\Phi_2$ contains terms of $O(\mathcal{P}^{-1})$. We have set

$$\lambda_\perp = -S \cdot p_\perp/\mu \tag{A.25}$$

and $\mu$ is an undetermined energy scale, which we shall fix below. Moreover the distributions involved, for which we have used the notations of MT, are functions of the Bjorken variable $x$ and of the intrinsic transverse momentum $\mathbf{p}_\perp$. The term

$$\Phi_0 = \Phi_{0a} + \Phi_{0b} \tag{A.26}$$

is interaction independent. This is evident for $\Phi_{0a}$, which consists of twist-2 operators. We shall show that also $\Phi_{0b}$ shares this feature, although the operators involved are classified as "twist-3".

### A.3 - Relationship between $h_{1T}$ and $g_{1T}$

*A.3.1 - Comparison with the QCD parton model*

We equate the coefficients of the independent Dirac operators in eqs. (A.17) and (A.21), taking into account the first eq. (A.2), which in the $\mathcal{P}$-frame reads

$$p = \sqrt{2} x \mathcal{P} n_+ + p_\perp + O\left(\mathcal{P}^{-1}\right). \tag{A.27}$$

We get

$$f_1 = f_1^\perp = q, \tag{A.28}$$

$$\lambda_\perp h_T^\perp = sin\theta' \delta q_\perp, \tag{A.29}$$

$$\lambda_\perp h_{1T}^\perp = (1-\epsilon_1) sin\theta' \delta q_\perp, \tag{A.30}$$

$$\mu\lambda_\perp h_T = (1-\epsilon_1)sin\theta' E_q\delta q_\perp. \tag{A.31}$$

$$\lambda_\perp g_{1T} = (1-\epsilon_2)sin\theta'\delta q_\perp, \tag{A.32}$$

$$\lambda_\perp g_T^\perp = (1-\epsilon_3)sin\theta'\delta q_\perp. \tag{A.33}$$

Here $\epsilon_1 = m/E_q$, $\epsilon_2 = m/2x\mathcal{P}$ and $\epsilon_3 = m/2|\mathbf{p}|$ are the correction terms to the chiral limit, which are generally small for light quarks. The terms of order $O\left[(m^2 + \mathbf{p}_\perp^2)/\mathcal{P}^2\right]$ have been neglected. As regards $\mu$, RS and MT have set it equal to $M$. We require the various functions to be normalized, in the chiral limit, as $\delta q_\perp$, which is a difference between two probability densities. Therefore we assume

$$\lambda_\perp = sin\theta', \tag{A.34}$$

which, according to eqs. (A.10) and (A.25), implies $\mu = |\mathbf{p}|$. In particular, this choice leads to the relationship

$$g_{1T} = (1-\epsilon_2)h_{1T}. \tag{A.35}$$

In the chiral limit, $sin\theta' h_{1T} = sin\theta' g_{1T}$ is the average helicity of a quark in a transversely polarized proton.

*A.3.2 - Equations of motion*

Owing to the Politzer theorem[37], the Fourier transform of $\Phi_\perp$ must fulfil the equation of motion (e.o.m.) for a Dirac particle interacting with the gluon field. We set $\Phi_\perp = \Phi_\perp^{free} + \Phi_\perp^{int}$. Since the term $\Phi_\perp^{free}$ (eq. (A.8)) fulfils the Dirac equation for a plane wave, the e.o.m. implies that $\Phi_\perp^{int}$ depends on the quark-gluon interaction and is of order $g\mathcal{P}^{-1}$[28, 56], $g$ being the strong interaction coupling constant. But $\Phi_\perp^{free}$ includes the term $\tilde{\Phi}_0$, corresponding to eq. (A.26) with the constraints (A.28)-(A.33). On the other hand, $\Phi_\perp - \tilde{\Phi}_0$ includes the interaction dependent term $\Phi_\perp^{int}$ and is orthogonal to $\tilde{\Phi}_0$: $Tr\left[\tilde{\Phi}_0(\Phi_\perp - \tilde{\Phi}_0)\right] = 0$. Therefore $\tilde{\Phi}_0$ is interaction independent and the relationships (A.28) to (A.33), although deduced from the naive parton model[23], hold true even after inserting interactions. In particular the term $\Phi_{0b}$, although made up with twist-3 operators, is interaction independent and should be classified as a "kinematic" higher twist term.

The Politzer theorem survives renormalization and off-shell effects[37], which, on the other hand, do not cause any mixing among the mutually orthogonal Dirac operators into which we have decomposed the correlation matrix. Therefore the relationships (A.28) to (A.33) have a quite general validity; in particular they hold also when QCD evolution is taken into account.

*A.3.3 - An alternative proof of relationship (A.35)*

Eqs. (A.21) and (A.1) imply

$$x\mathcal{P}\lambda_\perp g_{1T} = -\frac{1}{2}Tr\left[\Phi\gamma_5\slashed{n}_-\right] = -\frac{1}{2}\mathcal{H}\left[\langle P,S|\overline{\psi}(0)\gamma_5\slashed{n}_-\psi(y)|P,S\rangle\right], \qquad (A.36)$$

having defined the functional

$$\mathcal{H}\left[\rho(y)\right] = \int dp^- \int \frac{d^4y}{(2\pi)^4}e^{ipy}\rho(y). \qquad (A.37)$$

Similarly,

$$x\mathcal{P}h_{1T} = \frac{1}{4}Tr\left\{\Phi_0\gamma_5[\slashed{S},\slashed{n}_-]\right\} = \frac{1}{4}\mathcal{H}\left[\langle P,S|\overline{\psi}(0)\gamma_5[\slashed{S},\slashed{n}_-]\psi(y)|P,S\rangle\right]. \qquad (A.38)$$

Now we decompose the quark field into the eigenstates of a canonical representation, in which the quantization axis is taken along the proton polarization:

$$\psi(y) = \psi'_\uparrow(y) + \psi'_\downarrow(y) + \psi''(y). \qquad (A.39)$$

Here $\psi'$ and $\psi''$ denote respectively the "good" and "bad" component of the quark field, *i. e.*,

$$\psi'_{\uparrow(\downarrow)}(y) = \frac{1}{4}(1\pm\gamma_5\slashed{S})\slashed{n}_+\slashed{n}_-\psi(y), \qquad \psi''(y) = \frac{1}{2}\slashed{n}_-\slashed{n}_+\psi(y). \qquad (A.40)$$

The field may also be decomposed into chirality eigenstates, *i. e.*,

$$\psi(y) = \psi'_R(y) + \psi'_L(y) + \psi''(y), \qquad \psi'_{R(L)}(y) = \frac{1}{4}(1\pm\gamma_5)\slashed{n}_+\slashed{n}_-\psi(y). \qquad (A.41)$$

Then

$$\overline{\psi}(0)\gamma_5\slashed{n}_-\psi(y) = -\frac{1}{\sqrt{2}}\left[\psi'^\dagger_R(0)\psi'_R(y) - \psi'^\dagger_L(0)\psi'_L(y)\right], \qquad (A.42)$$

$$\frac{1}{2}\overline{\psi}(0)\gamma_5[\slashed{S},\slashed{n}_-]\psi(y) = +\frac{1}{\sqrt{2}}\left[\psi'^\dagger_\uparrow(0)\psi'_\uparrow(y) - \psi'^\dagger_\downarrow(0)\psi'_\downarrow(y)\right]. \qquad (A.43)$$

Substituting eqs. (A.42) and (A.43) respectively into (A.36) and (A.38), we get

$$\lambda_\perp g_{1T}(x, p_\perp) = q_R^\Uparrow(x, p_\perp) - q_L^\Uparrow(x, p_\perp), \tag{A.44}$$

$$h_{1T}(x, p_\perp) = q_\uparrow^\Uparrow(x, p_\perp) - q_\downarrow^\Uparrow(x, p_\perp). \tag{A.45}$$

Here we have set

$$q_R^\Uparrow = \frac{1}{2\sqrt{2}x\mathcal{P}} \mathcal{H}\left[\langle P, S|\psi_R'^\dagger(0)\psi_R'(y)|P, S\rangle\right], \tag{A.46}$$

$$q_\uparrow^\Uparrow = \frac{1}{2\sqrt{2}x\mathcal{P}} \mathcal{H}\left[\langle P, S|\psi_\uparrow'^\dagger(0)\psi_\uparrow'(y)|P, S\rangle\right] \tag{A.47}$$

and we have defined analogously $q_L^\Uparrow$ and $q_\downarrow^\Uparrow$. Eqs. (A.46) and (A.47) imply that $q_{R(L)}^\Uparrow$ is the probability of finding the quark in a chirality state, whereas $q_{\uparrow(\downarrow)}^\Uparrow$ is the probability for a quark to be in a state of the canonical represention defined above. In this representation the spin density matrix of a quark in a transversely polarized proton reads

$$\rho = q_\uparrow^\Uparrow |\uparrow\rangle\langle\uparrow| + q_\downarrow^\Uparrow |\downarrow\rangle\langle\downarrow|, \tag{A.48}$$

$|\uparrow(\downarrow)\rangle$ denoting the two states of the canonical representation. On the other hand, the helicity operator can be written as

$$\Lambda = |+\rangle\langle+| - |-\rangle\langle-|, \tag{A.49}$$

where $|\pm\rangle$ denote the two helicity states. Therefore the average helicity of a quark in a transversely polarized proton results in

$$\begin{aligned}\langle\lambda_\Uparrow\rangle &= Tr(\rho\Lambda) = q_\uparrow^\Uparrow \left[|\langle+|\uparrow\rangle|^2 - |\langle-|\uparrow\rangle|^2\right] + q_\downarrow^\Uparrow \left[|\langle+|\downarrow\rangle|^2 - |\langle-|\downarrow\rangle|^2\right]\\ &= \left(q_\uparrow^\Uparrow - q_\downarrow^\Uparrow\right) sin\theta' = h_{1T} sin\theta'.\end{aligned} \tag{A.50}$$

But $\langle\lambda_\Uparrow\rangle$ equals $q_R^\Uparrow - q_L^\Uparrow$ in the chiral limit. Then

$$q_R^\Uparrow - q_L^\Uparrow = (q_\uparrow^\Uparrow - q_\downarrow^\Uparrow) sin\theta'. \tag{A.51}$$

The definitions (A.46) and (A.47) and equations (A.42) and (A.43) imply

$$\mathcal{H}\left[\langle P, S|\overline{\psi}(0)O\psi(y)|P, S\rangle\right] = 0 \qquad \text{for} \qquad m = 0, \tag{A.52}$$

having set

$$O = \gamma_5 \left\{ \not{n}_- - \frac{1}{2} sin\theta'[\not{S}, \not{n}_-] \right\}. \tag{A.53}$$

Therefore, in the chiral limit, relationship (A.35) holds true independent of renormalization and off-shell effects.

### A.4 - Discussion

The result we have just found appears in contrast with the behavior of the ordinary distributions. Indeed, $g_1(x)$, a chiral even function, is independent of $h_1(x)$, which is chiral odd. Therefore an experiment for determining $g_1(x)$ automatically excludes the possibility of inferring $h_1(x)$ and *vice-versa*. But transverse momentum attenuates this difference. A quark polarized perpendicularly to the proton momentum has a nonzero helicity if its transverse momentum is different from zero. Quantitatively, in a transversely polarized proton, a massless quark, whose average spin component along the proton spin is $h_{1T}(x, \mathbf{p}_\perp)$, has an average helicity $sin\theta' h_{1T}(x, \mathbf{p}_\perp)$, which implies $h_{1T}(x, \mathbf{p}_\perp) = g_{1T}(x, \mathbf{p}_\perp)$ in the chiral limit. The fact that $h_{1T}$ is associated to a chiral odd operator and $g_{1T}$ to a chiral even one simply means that the same function can be deduced from different types of experiment: either from single polarization, exploiting the Collins effect[25], or from double polarization, as suggested in the present paper and by Kotzinian and Mulders[21].